\documentclass{elsart}

\usepackage{graphics}
\usepackage{graphicx}
\usepackage{color}
\usepackage{epsfig}

\usepackage{amssymb}
\DeclareGraphicsExtensions{.pdf,.jpg,.eps,.ps}
\usepackage{graphicx}
\usepackage{cite}
\usepackage{array}

\begin{document}

\begin{frontmatter}

% Title, authors and addresses

% use the thanksref command within \title, \author or \address 
% for footnotes;
% use the corauthref command within \author for corresponding author 
% footnotes;
% use the ead command for the email address,
% and the form \ead[url] for the home page:
% \title{Title\thanksref{label1}}
% \thanks[label1]{}
% \author{Name\corauthref{cor1}\thanksref{label2}}
% \ead{email address}
% \ead[url]{home page}
% \thanks[label2]{}
% \corauth[cor1]{}
% \address{Address\thanksref{label3}}
% \thanks[label3]{}

\title{A Vertex Trigger based on Cylindrical Multiwire Proportional Chambers
}

\author{J.~Becker\thanksref{sschmitt}},
\author{K.~B\"osiger},
\author{L.~Lindfeld},
\author{K.~M\"uller},
\author{P.~Robmann},
\author{S.~Schmitt\thanksref{sschmitt}},
\thanks[sschmitt]{Present address: DESY, D-22607 Hamburg, Germany}
\author{C.~Schmitz},
\author{S.~Steiner},
\author{U.~Straumann\corauthref{cor}},
\corauth[cor]{Corresponding author. Phone: +41~44~635~5768, 
Fax: +41~44~635~5704} \ead{strauman@physik.unizh.ch}
\author{K.~Szeker},
\author{P.~Tru\"ol},
\author{M.~Urban\thanksref{murban}},
\thanks[murban]{Present address: Philips Medical Systems DMC GmbH, 
D-22335 Hamburg, Germany}
\author{A.~Vollhardt},
\author{N.~Werner},
\address{Physik-Institut, Universit\"at Z\"urich, CH-8057 Z\"urich, 
Switzerland}

\author{D.~Baumeister\thanksref{dbaumeister}},
\thanks[dbaumeister]{Present address: Continental Teves AG \& Co., 
oHG, D-60488 Frankfurt, Germany}
\author{S.~L\"ochner\thanksref{sloechner}},
\thanks[sloechner]{Present address: GSI, D-64291 Darmstadt, Germany}
\address{ASIC--Laboratory, Kirchhoff--Institut f\"ur Physik, 
D-69120 Heidelberg, Germany} 

\author{M.~Hildebrandt},
\address{Paul Scherrer Institut, 5232 Villigen PSI, Switzerland}

\begin{abstract}
% Text of abstract
This article describes the technical implementation and the performance of the 
$z$-vertex trigger (CIP2k), which is part of the H1-experiment at HERA.

The HERA storage ring and collider was designed to investigate 
electron (and positron) proton scattering at a center-of-mass energy of 
320 G$e$V. To improve the sensitivity for detecting non-standard 
model physics and other high momentum transfer phenomena, the HERA ring has 
been ugraded between 2000 and 2003 to increase the specific luminosity for
the experiments. In order to cope with the increased event and background rate 
the experiments were upgraded, too. The CIP2k trigger system 
is based on a set of five cylindrical multiwire 
proportional chambers with cathode pad readout, 
and allows to distinguish between events 
induced by beam background and $ep$-interactions at the first trigger stage. 
The trigger decision is calculated dead-time free with a latency of 
1.5\,$\mu$s in parallel to the beam clock at 10.4\,MHz. 
The trigger-logic is realized in large field programmable gate arrays 
(FPGA) using the hardware description language Verilog.
The system is operational since October 2003. It suppresses background 
events with high efficiency and provides event timing information, 
as designed.
\end{abstract}

\end{frontmatter}

% Running line numbers:
%\pagewiselinenumbers

% ****************************************************************
 % intro
% \input{nimintroduction}
% main text
\section{Introduction}\label{sec:intro}

%upgrade: pros
During the period 1993-2000 each of the two experiments at the HERA collider, 
H1 and ZEUS, collected an integrated luminosity of 
approximatly 100$\,\mathrm{pb}^{-1}$ (HERA~I run). The subsequent upgrade 
of the accelerator lead to a significantly increased specific luminosity in
2004, namely $1.7\times 10^{30} $cm$^{-2}$s$^{-1}$mA$^{-2}$ 
three times higher as compared to HERA~I. 
The target for the integrated luminosity is $250\,\mathrm{pb}^{-1}$ per year.
This was achieved by increasing the proton beam current and 
by reducing the beam cross sections at the interaction point (IP).
The latter required the installation of strong superconducting 
focussing magnets inside the H1 experiment~\cite{h1det} 
close to the IP, which allow 
an early separation of the electron and the proton beam.
The new beam parameters are listed in Table~\ref{hera2} in comparison 
to those of HERA~I. 
\begin{table}[ht]
\centering
\begin {tabular} {|l l||c|c|}
\hline 
 & & HERA~I &  HERA~II\\
\hline
\hline   Electron (positron) current & [mA]  &   50 &   55 (35) \\
\hline   Proton current & [mA]  &   100 &   135 (105) \\
\hline   Interaction focus & [$\mu{\rm m}^2$]  &  $ 180\times 50$&  
$110\times 30$ \\
\hline   Specific luminosity & [cm$^{-2}$s$^{-1}$mA$^{-2}$]  &  
$ 5\times 10^{29}$&  $ 1.7\,(1.6)\,\times 10^{30}$\\
\hline
\end{tabular}
\caption{Comparison of HERA~II and HERA~I beam parameters. For HERA II
the design values and, in brackets, typical running conditions in autumn 
2006 are given.  \label{hera2}}
\end{table}

The change of the beam optics led to significantly increased beam 
induced background in the experiment, whose separation 
from $ep$ interaction as fast as possible is crucial for efficient 
data taking. In the H1 experiment the corresponding event-by-event trigger
decision is based on timing or on the reconstructed vertex position 
using fast drift or multiwire proportional chamber data. 
In HERA~I running a trigger based on the information
of two sets of two cylindrical multiwire proportional chambers 
(central inner (CIP) and central outer (COP) MWPC) 
was used for this purpose~\cite{oldtrigger,eic93}. It reconstructed the vertex 
position along the beam axis within $\pm$30 cm around the IP and 
delivered an event timing. 
This information was part of most H1 trigger decisions.

To cope with the high background rate the central inner proportional 
chambers of the H1 experiment and the trigger based on 
their signals were redesigned to allow reconstruction of the vertex 
position over a range of 3.5~m along the beam axis ($z$-vertex). 
With increased acceptance beam induced background can be 
more efficiently rejected. 

The construction of the new central inner proportional chamber, the design and 
the implementation of the $z$-vertex trigger derived from it, and
the performance of the complete system are described in this article.

\subsection{Background situation}\label{ssec:backg}
\begin{figure}[ht]
\centering
\includegraphics[width=0.98\textwidth]{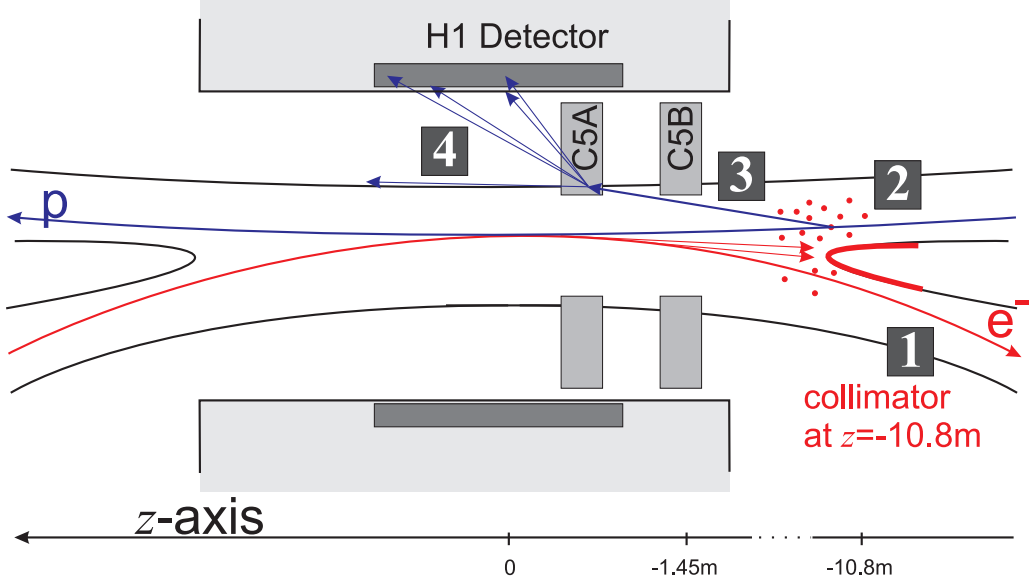}
\caption[Synchrotron Radiation Effects.]
{Synchrotron radiation background: 
Bending of the electrons (1) leads to a high amount of synchrotron radiation
which reduces the quality of the vacuum in the beam pipe (2). 
As a consequence the background rate rises due to collisions of protons 
with rest gas nuclei (3), which are observed in the H1 detector 
as hadronic showers (4)}
\label{syncradi}
\end{figure}

%upgrade: cons
The deflection of the electron beam induced by the strong focussing magnets 
near the interaction point creates a high level of synchrotron 
radiation of about 100\,kW intensity. The interaction region is designed such 
that most of this synchrotron radiation is absorbed far away from 
the experiment. 
The closest absorber for synchrotron radiation is located at $z=-10.8$ m, i.e. 
10.8 m away from the IP\footnote{At the H1 experiment, the $z$-axis
points along the proton flight direction, the $x$-axis points towards 
the center of the ring, the $y$-axis in
the upward direction. The origin of this right handed Cartesian 
coordinate system is the nominal interaction
point of the H1 experiment. The azimuthal angle in the $xy$-plane 
is $\varphi$. The polar angle is $\Theta$ (positive\,$z$, $\Theta=0$). 
The parts of the detector with positive (negative) $z$ coordinates are 
often referred to as the forward (backward)regions.}.
A small fraction of the synchrotron radiation is reflected back rather than 
being absorbed. 
Two collimators in the beam pipe (C5A at $z=-80$~cm and C5B at $z=-145$~cm)  
protect the H1 detector against this backscattered synchrotron radiation. 
The setup is depicted in Fig.\,\ref{syncradi}. 
The synchrotron radiation leads to increased 
rest-gas pressure in the beam pipe close to the absorber at $z=-10.8$ m, and
consequently the rate of collisions of protons with the gas molecules
is high. In these beam-gas interactions off-momentum particles are produced
creating hadronic showers through secondary interactions in the collimators,
and high particle multiplicities in the tracking detectors. 
At HERA~II this mechanism is the dominant source of beam induced 
background \cite{bkgreport}.

In order to use the high luminosity provided by HERA~II efficiently 
the fraction of background events has
to be suppressed at an early stage of the trigger. 
Due to the fixed bandwidth of the H1 data acquisition system, it is mandatory
to identify these interactions without causing any interruption of data taking.
Each event which is not filtered out at the first level 
of the pipelined trigger system causes deadtime.

\subsection{Background identification}\label{ssec:backgid}

Background events can be identified if their origin along the beam axis
($z$-vertex) is known. This is obvious from inspecting 
Fig.\,\ref{zvertex_cip} in Sec.~\ref{sssec:bins}, 
which shows the reconstructed 
$z$-vertex distribution observed in a typical HERA~II run. 
Beam induced background events originating 
from the collimators C5A and C5B and $ep$-interactions can clearly be
distinguished. 
The CIP2k $z$-vertex trigger system was therefore developed to 
reconstruct the trajectories of charged particles coming
from interactions near the  beam axis. 
It determines the vertices and creates a corresponding histogram for 
each bunch crossing. 
Event signatures with most of the tracks pointing 
to the backward region are rejected by the trigger.

The track reconstruction is based on data from  a set of five cylindrical
multiwire proportional chambers (central inner proportional chambers: CIP2k) 
with cathode pad readout. As the name indicates the CIP2k encompasses 
the interaction region in the center of H1 detector.
It has a total  of 8480 readout pads, providing space points and 
timing information. Each layer is equally subdivided into 16 azimuthal
segments in $\varphi$ with approximately 20 mm wide segmentation 
along the beam axis. 
The chamber has sufficient granularity to resolve the event structure, 
even in high multiplicity events. Five layers ensure a high redundancy for the
reconstruction with a negligible rate of combinatorial background.
 Operation of the trigger system remains efficient even if one or two layers
 of the chamber are not operational. 
Simulation studies were done beforehand in order to optimize the 
granularity of the chamber and to determine the expected
redundancy and background rejection power\cite{simulationsCIP,ciptheses}.
Since multiwire proportional chambers are characterized by a response 
time below 10\,ns \cite{sauli}, the
CIP2k $z$-vertex trigger delivers accurate event timing ($t_0$) 
with optimal efficiency. 

All H1 detectors and trigger systems store their information for 
32 bunch crossings (3.1\,$\mu$s)~\cite{h1det}. 
The level 1 (L1) trigger decision 
starts the readout after 2.3\,$\mu$s, 
hence all L1 signals have to be delivered not later than 2\,$\mu$s.
The decision has to be made for every bunch crossing. 
Thus all pads have to be read out within 96 ns, corresponding 
to the bunch crossing frequency of 10.4 MHz. In the CIP2k trigger system 
this requirement is met using 
large field programmable gate arays (FPGA) both for the trigger
decision and the storage of the event data for readout by the data 
acquisition system. 

% ****************************************************************
 % chamber
% \input{nimchamber}
% main text
% -------------------------------------
\section{Multiwire proportional chamber}\label{sec:chamber}
\subsection{Construction}\label{ssec:chconstr}
The construction of the CIP2k in the University of Z\"urich shops 
followed closely that of a thin cylindrical 
proportional chamber with cathode pad readout \cite{oldtrigger} 
which was installed in H1 during the HERA~I run.
This chamber also delivered signals for a $z$-vertex trigger \cite{eic93} 
and ran very stable  and efficiently over the whole data 
taking period of HERA~I.

\begin{figure}[ht]
\centering
\includegraphics[width=0.90\textwidth]{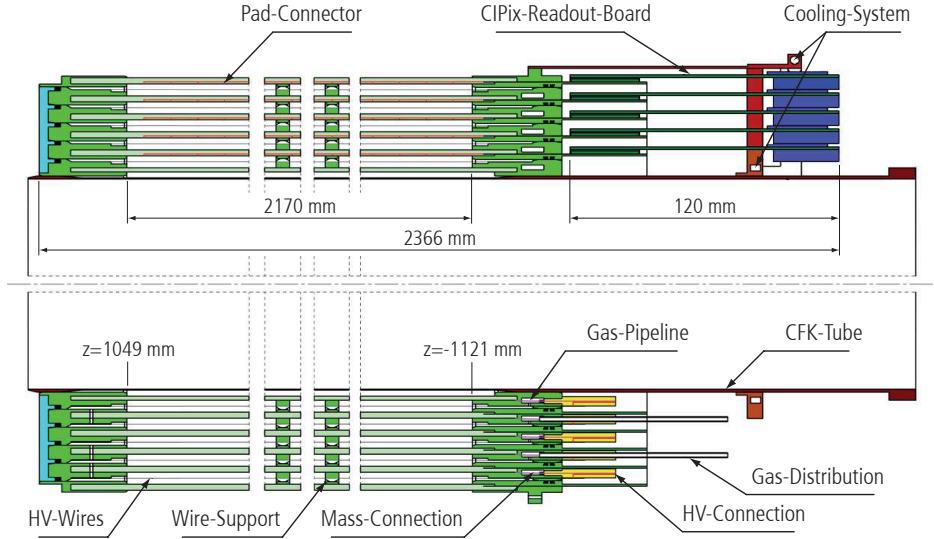}
\caption[Side view of the CIP2k chamber]
{Cross section of the five layer cylindrical multiwire proportional chamber 
with cathode pad readout.\label{fullchamber}} 
\end{figure}

The cylindrical chamber has five radial layers with an inner radius of
154\,mm and an outer radius of 196\,mm and an active length of 2170 mm. 
The total length of the chamber 2366 mm including end flanges and 120 mm for
the electronics and the cooling. The chamber 
encloses the nominal interaction region and is mounted between 
the central silicon detector~\cite{h1silicon} on the inside and 
the central drift chamber (CJC)~\cite{h1det} on the outside. 
The active region extends from $z$=-1121\,mm 
to $z$=1049\,mm, corresponding to an angular acceptance  
$11^\circ<\Theta<169^\circ$. Figure \ref{fullchamber} shows a 
cross section of the chamber. All necessary details for the construction
techniques and the choice of materials are given in ref.~\cite{steiner}. 

%construction method + principle

\begin{figure}
\centering
\includegraphics[width=0.90\textwidth]{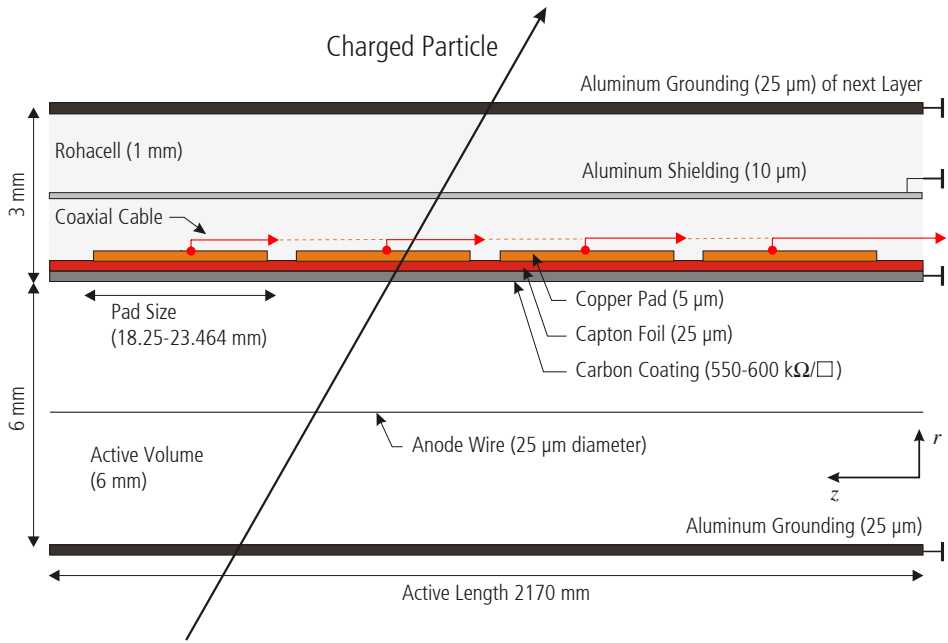}
\caption[Side view of the CIP2k chamber]
{Side view of one layer of the CIP2k chamber in the $rz$-plane.
%The anode and induced cathode pad signals are indicated.
\label{scheme_cathodepad}}
\end{figure}

The sandwich-like construction of each radial layer shown in 
Fig.\,\ref{scheme_cathodepad} was chosen in order to keep the material
budget low without compromizing on stability and robustness.
Each layer contains 480 gold-plated tungsten 
anode wires, $\approx$2\,mm spaced, 
with a diameter of 25\,$\mu$m and mounted in a gas volume of $\approx$ 16 
liters. They are kept at a voltage between 2250 and 2500 V. At the outer
layers a lower voltage is needed, since electric field on the wire is higher.
The wires are supported by epoxy rings at one and two thirds of the chamber
length. 
A 3 mm gap separates the anode from the cathode planes, which consist an 
aluminum foil (25\,$\mu$m thickness) on the inner and a carbon coated capton 
foil on the outer side. The charge collected on the anode wires induces 
a charge on the high resistivity cathode planes (550-600\,k$\Omega /\square$) 
and a signal on the copper pads on the outer side of the capton foil. 
Copper was chosen as pad material instead of 
aluminum as in the original central inner proportional 
chamber~\cite{oldtrigger}, because a large fraction of the induced signal was
lost in the latter case where the pad thickness was smaller than the skin
depth. Furthermore the signal transmission was improved by using
miniature coaxial cables~\cite{coax,kollak} with very low dielectric and 
inductive 
coefficients to transport the induced charge (${\bf O}(10^5)$ \,electrons) 
of each pad to a connector at the rear end flange of the 
chamber. These cables 
are embedded in groves in the layer of Rohacell~\cite{rohacell} (thickness
1mm), which backs the foils and guarantees for the mechanical stability.
Each pad covers 22.5$^\circ$ in azimuth. In $z$-direction, the size and number 
of pads vary at each layer as described below.   

\subsection {Pad geometry}\label{ssc:pads}

The arrangement of the pads is optimized for a simple trigger algorithm. 
While the pad size is constant within each layer 
along $z$, it increases proportional to the
distance from the beam axis from 18.25 to 23.464 mm ({\it projective
  geometry}). The innermost pad size
was dictated by the requirement that two pads taken in a logical OR match the
granularity of the previous set up~\cite{eic93}.
The pads are arranged such that tracks from the same vertex generate 
similar patterns independent of $\Theta$, as described in 
Sec.\,\ref{ssec:trigalg}. Every 328.50 mm the pads of the five layers 
are exactly aligned above each other. This distance equals two bins in the
$z$-vertex histogram. In total, there are 8,480 pads in 16\,$\varphi$-sectors
(see Table\,\ref{chamberdata}). 

\begin{table}[ht]  
\centering
\begin {tabular} {|c||c|c|c|c|c|}
\hline Layer &  0 & 1 & 2& 3& 4\\
\hline
number of pads in $z$ & 119 & 112 & 106 & 99 & 93 \\
total number of pads & 1904 & 1792 &1696 & 1584 & 1488 \\
length of pad [mm] & 18.250 & 19.323 & 20.531 & 21.900 & 23.464 \\
length of last pad [mm] & 16.0 & 19.323 & 13.74 & 21.900 & 10.8 \\
anode radius [mm] & 157 & 166 &  175 & 184 & 193 \\
spacing of wires [mm] & 2.06 & 2.17 & 2.29 & 2.41 & 2.53 \\
\hline
\end{tabular} 
\caption{
Geometry of readout pads and position of the anode wires.
 The last pads ($+z$ side) in layers\,0, 2 and 4 have a reduced length.  
\label{chamberdata}}
\end{table}

\subsection{Gas system and high voltage distribution}\label{ssec:gashv}
The chamber is operated with a gas mixture of argon ({49.9\%}), 
isobutane ({49.9\%}) and freon ({0.2\%}). 
The gas flow is controlled by the H1 wide gas control system~\cite{h1det}.
Two independent closed loops supply the five layers. 
The static pressure of the chambers relative to atmospheric pressure is 
regulated  to be within $\pm~ 50\, \mu$bar. 
This ensures that the active gap in the chamber is not distorted.
 Since the tracking detectors are surrounded by a N$_2$ atmosphere, 
contamination with N$_2$ through 
leakage and other changes in the gas mixture are monitored constantly in 
order to avoid changes in the gas amplification.

\begin{figure}[ht]
\centering
\includegraphics[width=0.70\textwidth]{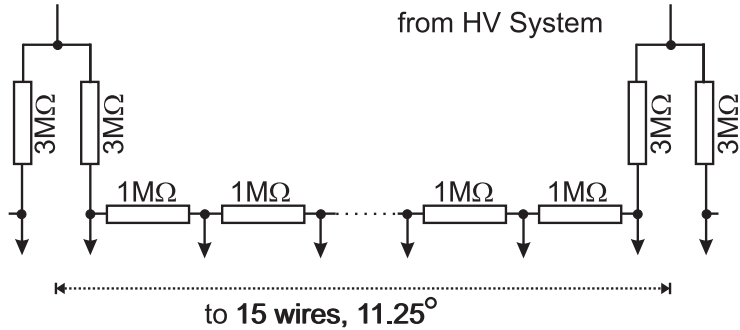}
\caption{
Two HV supplies~\cite{caen} with trip protection are connected 
to 15 HV wires of the chamber covering 11.25$^\circ$ in $\varphi$.}
\label{HV_system}
\end{figure}

The 480 anode wires of each layer are grouped into 32 sectors of 15 wires. 
Each group is connected to a high voltage cable via 3.3~M$\Omega$ resistor,
while the anode wires are connected to their neighbours with a 
1~M$\Omega$ resistor. This allows a high voltage degrading in case 
of a short circuit in one region of the chamber  with minimal efficiency 
loss~\cite{hvwerner}. An illustration of the high voltage distribution 
is given in Fig.\,\ref{HV_system}.

\subsection{Front end electronics}\label{ssec:frontelec}
The pad information is transferred to the front end electronics via 
a 120-pin connector on the chamber. 
Sixteen connectors per layer  transmit the 
signals of all pads along $z$ of one $\varphi$-sector.
The front end electronics consists of 80 highly integrated readout boards 
with a size of $5\times 13\,\mathrm{cm}^{2}$ in 12-layer printed circuit 
board (PCB) technology.
They were developed at the Paul Scherrer Institut.
Each board covers one $\varphi$-sector and contains two 
 application specific integrated circuits (ASIC) especially developed for the 
readout (called CIPix) and two 17:1 multiplexers~\cite{hp1032}.
The information of one double layer is sent via an optical fibre to the 
receiver cards 40 m away in the electronics trailer of the experiment.

\subsubsection{The CIPix readout chip}\label{sssec:cipix} 

\begin{figure}[ht]
\centering
\includegraphics[width=0.90\textwidth]{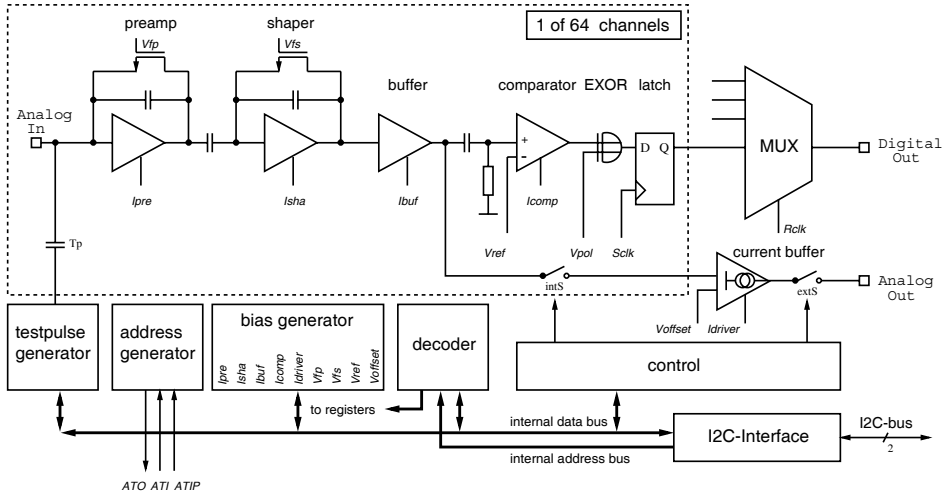}
\caption[Block diagram of the CIPix PCB]
{Block diagram of the CIPix PCB for one of the 64 channels. 
\label{CIPix_block}}
\end{figure}

Each CIPix readout chip has capacity for 64\,analog 
inputs, of which up to 60 are used while the remaining serve testing 
purposes. The CIPix readout chip was developed at the ASIC laboratory 
of the University of Heidelberg in a 0.8\,$\mu${m} CMOS-process. 
Figure\,\ref{CIPix_block} shows the block diagram of one CIPix chip.
Each of the 64 analog input channels consists of:
\begin{itemize}
\item A charge-sensitive preamplifier with a gain of 20\,mV per 
$10^5$ electrons. The preamplifier 
characteristics are programmable.
\item An analog signal shaping to a {\it CR-RC} semi-Gaussian pulse 
with program\-mable shaping times.
\item Digitalization by a one-bit comparator with programmable 
threshold and polarity. The digitized information 
is synchronized to the 10.4\,MHz beam clock signal.
\end{itemize}
   
The digitized pad information is multiplexed fourfold into 17 
digital channels (15 outputs and two control 
channels) at 41.6\,MHz in each CIPix. In parallel, one analog output 
channel is provided, which can be used to 
monitor any of the 64 input channels.
The first of the four multiplexed pads is tagged, and hence allows a 
synchronization with the demultiplexing electronics in the trigger system. 
All boards of one layer are interconnected by an $I^2C$ bus daisy-chain.
This allows to set all programmable parameters layer-wise and 
a serialized temperature measurement~\cite{da_vollhardt}. 
The $I^2C$ system itself is controlled 
by the experiment control system PVSS II \cite{ecssystem,pvss}.

\subsubsection{Readout board, optical link and cooling}\label{sssec:readout} 

\begin{figure}[ht]
\centering
\includegraphics[width=0.98\textwidth]{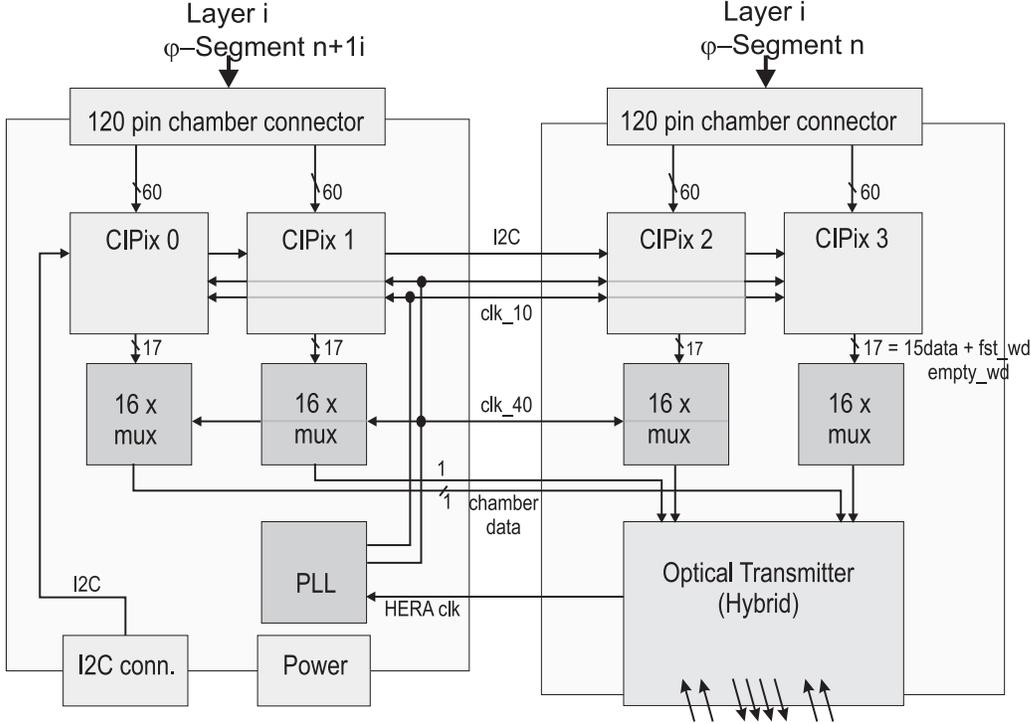}
\caption[A view of the CIPix double board]
{A schematic overview of the CIPix double board which reads out the pad 
information of two $\varphi$ sectors of one layer.
The two boards are interconnected by a flexible strip-line on a 
capton foil carrier to a double-board.
 One of the two boards holds an optical transmitter unit to transmit the 
information of one 
double-board with optical fibres  to the receiver cards. }
\label{PCB_scheme}
\end{figure}

\begin{figure}[ht]
\centering
\includegraphics[width=0.98\textwidth]{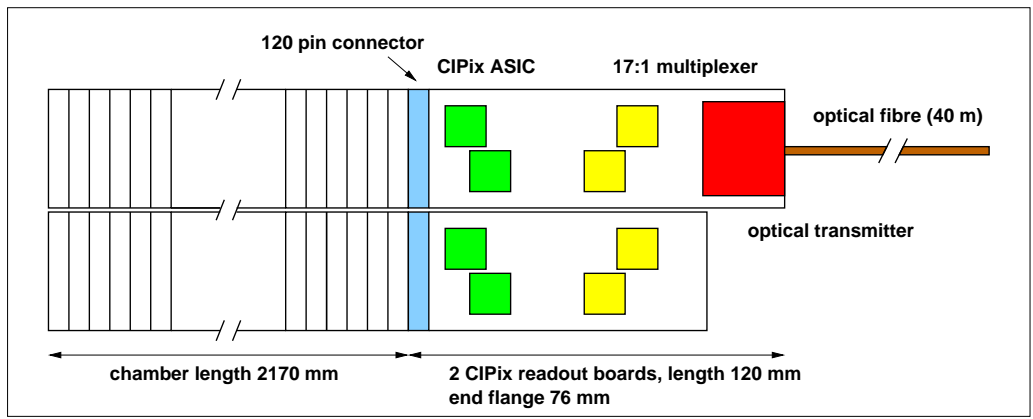}
\caption{Sketch of the CIP2k front end read out electronics.}
\label{syst_scheme_chamber}
\end{figure}

Figure\,\ref{PCB_scheme} provides an overview 
of the electronics on the chamber, while Figure~\ref{syst_scheme_chamber}
contains details of the front end PCB. 
The 15 plus 2 output channels of a CIPix are multiplexed 
to one output signal per CIPix and transmitted to an optical link 
system~\cite{luders_nim}. Pairs of adjacent boards are interconnected 
by a flexible strip-line on a capton foil carrier. 
One of the PCBs holds the optical link system while the other one 
holds the $I^2C$ interface and the voltage regulators. 
They are powered from the electronics trailer.
 The overall synchronisation is done with the global HERA clock signal.
It is the input for a low jitter phase-locked loop which
 generates the 41.6\,MHz clock. 
The latter  is  delivered to the 17:1 multiplexers and in addition to 
the HERA clock to  all four CIPix chips.

Since only optical transmission can support the high rates
with low cross talk and a minimal power consumption, an  
optical link between the on-detector electronics and the receiver cards 
was developed by ETH Z\"urich based on optical hybrids which do the 
electro-optical conversion~\cite{luders_nim}.
These hybrids have six sender and two receiver diodes for the
 10.4 MHz HERA clock. For redundancy two HERA clock signals are 
transmitted in parallel.
 The six senders transfer four digital data channels - one for each CIPix - 
with the signals of two $\varphi$ sectors and two analog channels - each 
shared by two CIPix - to the readout system.  
All the light guides from one transmitter unit are bundled into a single 
cable. 

The front end boards are cooled by direct contact to copper blocks 
which are connected to an outer and an inner cooling ring above 
and below the readout electronics. Both rings are water-cooled. 
The cooling of the boards is critical, because excessive heating may 
destroy the bond wires, which connect the readout ASIC to the board. 
Moreover, the connections within the 12\,layer PCB could also suffer.

% ****************************************************************
 % system description
 %\input{nimsystem}
\section{Trigger}\label{sec:trigger}

The CIP2k trigger electronics  serves two purposes:
\begin{itemize}
\item It uses the chamber information of all 8,480 pads to deliver 
trigger information for each 
bunch crossing to the H1 central trigger.
\item It stores all the chamber information in a pipeline and delivers 
it to the H1 
storage system in case of a positive trigger decision.
\end{itemize}

\subsection{Trigger algorithm}\label{ssec:trigalg}

A pipelined trigger algorithm has been developed which
is divided into five steps~\cite{phdurban}: 
\begin{enumerate}
\item Trackfinding
\item Assembly of the $z$-vertex histogram 
\item Histogram analysis 
\item Collection of information from the $\varphi$-sectors.
\item Trigger decision
\end{enumerate}
The first three steps are
done in parallel for every $\varphi$-sector in the trigger cards, 
the summation in step four in the sum cards, and the final trigger decision 
in the master sum card. The algorithm has been implemented in APEX 20k400 
FPGAs~\cite{fpga}, is written in the hardware description language Verilog
and made use of the development environment Altera Quartus (see ref. 
\cite{phdurban} for details).

\subsubsection{Track reconstruction:}\label{sssec:trackrecon}

\begin{figure}[h]
\centering
\includegraphics[width=0.98\textwidth]{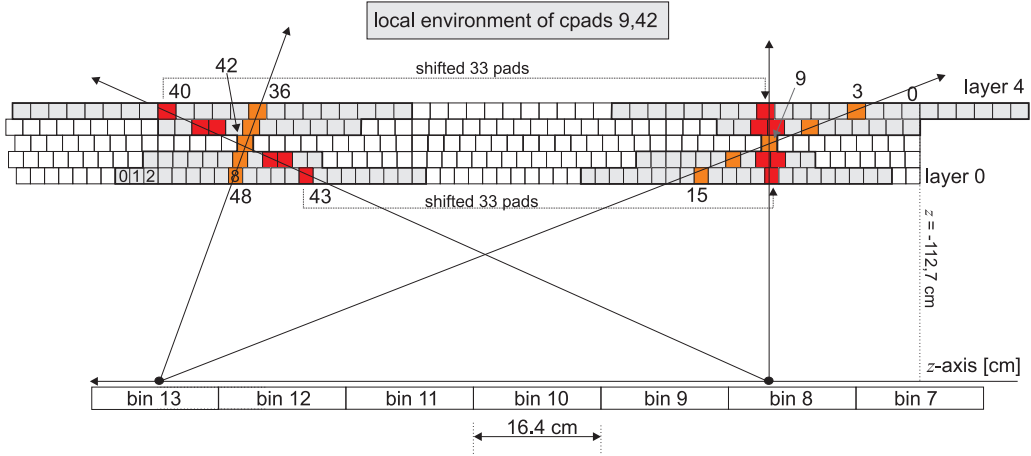}
\caption[Trackfinding with the projective chamber]
{Trackfinding with projective pad geometry: tracks from the
same vertex cause similar patterns in the local environment
of each central pad in layer 2. The grey shaded area shows the 
LE of central pad\,42 (left) and 9 (right), respectively. 
The track pattern 
pointing to $z$-bin\,13 is
defined in the same way for every LE: Pad\,8 in layer\,0, pad\,6 in
layer\,1, pad\,5 in layer\,3 and pad\,13 in layer\,4 define a track f
rom $z$-bin 13, similarly for track pattern 8 (13,9,3,9).}
\label{atrackfinding}
\end{figure}

The information coming from the receiver cards is fed  
into the FPGAs of the trigger where it is first demultiplexed four times.
Pads known to be defect may be permanently switched off or on in a filter 
module. At every
startup of the trigger system, a list of defect pads is updated via 
the VME bus.
In a second step valid track patterns with hits in at least four chamber
layers are searched for. The {\it projective geometry} of the chamber 
simplifies this task, since a given $z$-position leads to a unique pattern
of pads showing hits in the five layers. This is illustrated in 
Fig.~\ref{atrackfinding}. To allow for local inefficiencies the 
corresponding $z$-position of the track is stored even if a pad 
in a pattern is missing. The valid patterns are predefined in the 
{\it local environment} (LE) of every pad in the central layer.
Each possible track pattern in the LE then points to a given $z$-bin.
Since 22 $z$-bins are defined there exist 22 valid patterns in each LE.

\subsubsection{Assembly of the $z$-vertex histogram}\label{sssec:counting}

For each $z$-bin at total of 106 tracks can be identified
corresponding to the number of pads in layer\,2 and hence to 
the number of LEs.
If a track is identified in the trigger module it is written 
into a 106\,bit wide hit vector, defined for 
each of the 22 $z$-bins.
The tracks in the hit vector are counted for 
each bin in the adder module, which 
uses a major part of the FPGA logic because $22 \cdot 106 = 2332$ possible 
tracks have to be checked and summed up. 
The result is evaluated in less then two bunch crossings ($\leq 192$\,ns)
in a cascade of adders, which counts the tracks in the hit vector in six 
steps (144\,ns). 
Only 15 consecutive bins out of 22  are used for the trigger decision,
corresponding to 1590 possible tracks. 
They represent the $z$-vertex histogram for one 
$\varphi$-sector, calculated in one trigger
card. 
The selection of the 15 out of 22\,bins is programmable in the range 
$-235.4<z<+125.4$\,cm (see Fig.\,\ref{zvertex_cip}). 
For monitoring purposes, the histogram is stored in a pipeline and can 
be read out via the VME bus
together with the chamber information of the event.

\subsubsection{Histogram analysis}\label{sssec:bins}
\begin{figure}[h]  
\centering
\includegraphics[width=0.98\textwidth]{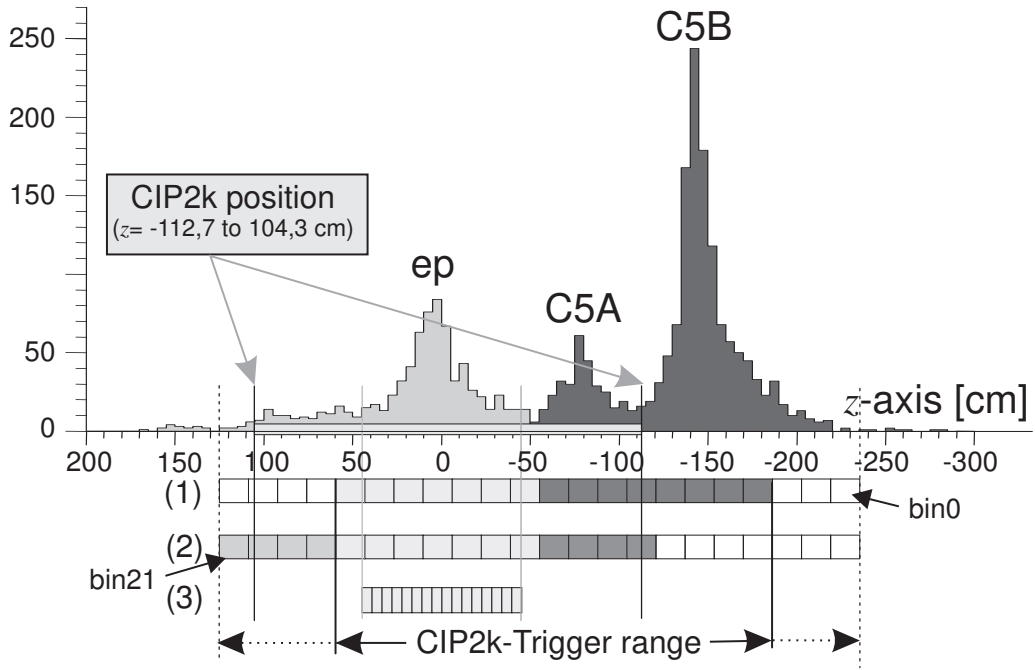}
\caption[$z$-vertex distribution and programmable regions of the 
$z$-vertex trigger]
{Reconstructed $z$-vertex distribution along the beam axis for a 
typical HERA~II run. Peaks appear at the nominal vertex position ($0\pm 20$
cm) and at the position of the collimators C5A ($-80$ cm) and C5B
($-145$ cm). Events which should be rejected by the $z$-vertex trigger 
are shaded in dark gray. The programmable regions of the CIP2k 
$z$-vertex trigger are shown below the histogram. 15 of 22 bins are 
available for the trigger. Configuration (1) is used to separate
background from $ep$ in luminosity runs, while configuration (2) is used 
for cosmic muon detection. The third line indicates the acceptance and the bin 
width of the old $z$-vertex trigger~\cite{eic93}.}
\label{zvertex_cip}
\end{figure}

For the trigger decision the bins of the $z$-vertex histogram are 
grouped into a forward, a central  and
a backward region. 
This selection is programmable as shown in Fig.\,\ref{zvertex_cip}
for two examples. The configuration adapted to $ep$ event 
fixes the backward region 
of the trigger to the range $-186 < z <-55$ cm and the central
region to  $-55<z<60$ cm. The second example concerns the cosmic muon trigger
which requires coincidences between two $\varphi$ quadrants. 

\subsubsection{Summing}\label{sssec:sum}
 Once the number of tracks in the three regions of each
$\varphi$-sector is known, they are added in two consecutive steps 
to give a global, $\varphi$ independent track distribution.
The information of four $\varphi$-sectors is added first 
in the pre-sum card to yield
$z$-vertex regions for the four quadrants of the CIP2k chamber. 
The number of tracks of the
quadrants is then added to get the global $z$-regions in the main sum card. 
To keep the number of transfer cables to a reasonably low 
level, the number of tracks is limited
to 255 in each region.

\subsubsection{Trigger decision}\label{sssec:decision}

A trigger decision is made by analyzing the track information in the 
forward, central and backward region based
on global track information of each event in the pipeline.
A precise timing information is evaluated (t$_0$, 1\,bit). It is set, if 
at least one track in the central
region is identified by the $z$-vertex trigger.
To analyze the quality of the $ep$-event, the number of tracks identified 
in all three regions (4\,bits) and the ratio between the number of central 
tracks and backward tracks (2\,bits) is given to the central H1 trigger
control. Up to 16 trigger elements (bits) \cite{phdurban} may be defined. 
They are transferred to the central
trigger system for every event.

\subsection{Trigger electronics}\label{ssec:trigelec}

\begin{figure}[ht]
\centering
\includegraphics[width=0.90\textwidth]{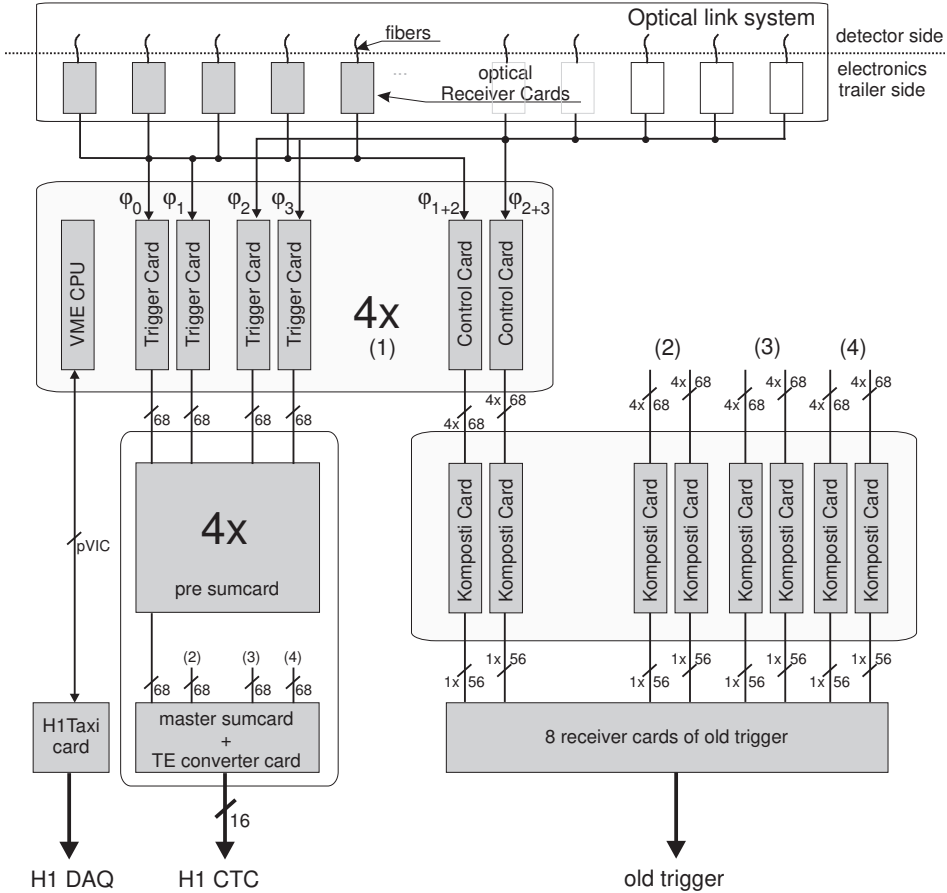}
\caption{
Overview of the CIP2k chamber and front end read out electronics. 
}
\label{syst_scheme_trailer}
\end{figure}

Seven crates house the trigger hardware: four trigger crates (TCr), one 
sum (SCr), one {\em komposti} ($\equiv$ {\em internal label}) and 
one subsystem trigger control (STC) crate
\cite{ieeeurban}. Figure\,\ref{syst_scheme_trailer} 
%and Table~\ref{numbers} 
provides an overview of the trigger electronics.

\subsubsection{Trigger crates:}\label{sssec:crates}

Each trigger crate hosts ten receiver (RC), four 
trigger (TC) and two control cards (CC) as well as 
one VME CPU. The crates have a special backplane and 
additional power supply units for 3.3\,V and 
-5\,V~\cite{wiener}.  

Each optical fiber originating from the CIPix PCB 
ends in a receiver card, which collects 
the information of four CIPix units, converts the optical signals 
to electrical signals and performs the 1:17 demultiplexing step. A clock
signal, phase-locked to the HERA clock, is also generated here. The pad and 
clock signals are sent to the P2-backplane of the trigger 
crate~\cite{luders_nim}. This backplane was specially developed at the
University of Heidelberg~\cite{ewhd} to cope with the 
high number of chamber signals. 
Ten 96-pin three-row connectors are needed. 
The signals of a group of five RCs are linked into two trigger cards 
on the front side. Each trigger card deals with the information 
of all five layers of one $\varphi$-sector. A control card serves
each pair of trigger cards and the five receiver cards linked to it with
the control signals and extracts the pad information of the two innermost 
layers to the old $z$-vertex trigger system~\cite{eic93}. 
The control card also  contains hardware to monitor and 
control the relative phases of the clock signals. 

\begin{figure}[h]  
\centering
\includegraphics[width=0.98\textwidth]{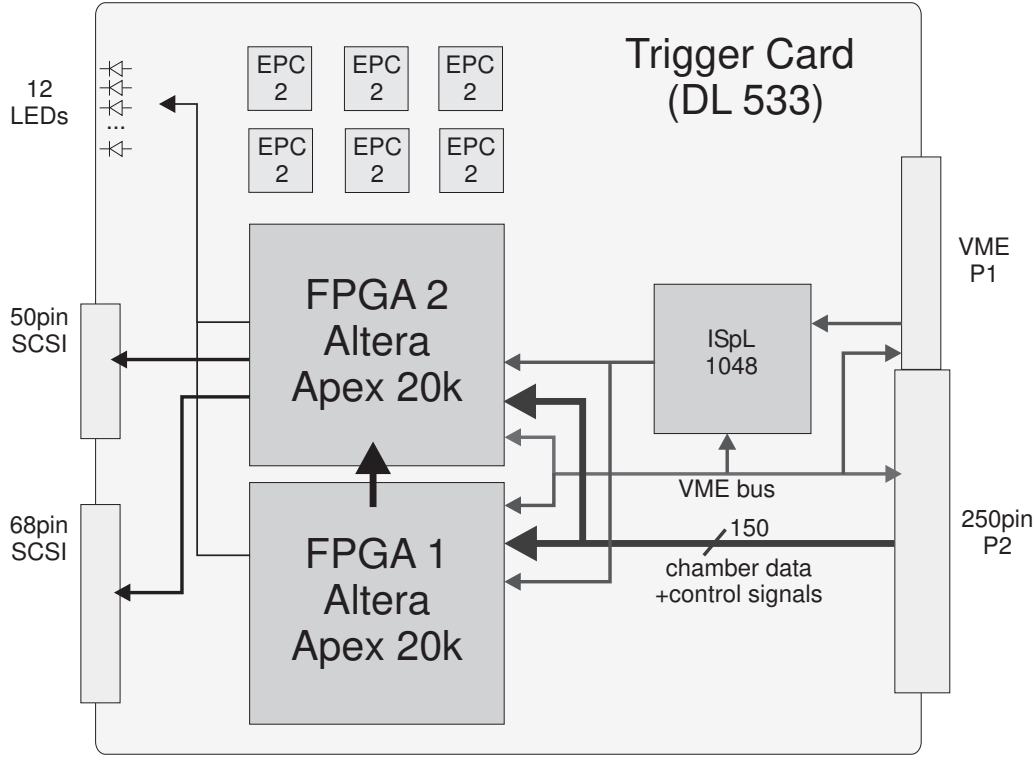}
\caption[The trigger card]
{
Schematic view of the trigger card (TC). 
}\label{triggercard}
\end{figure}

The trigger card contains both the logic of the trigger 
algorithm and the storage pipeline for the chamber data of one 
$\varphi$-sector. In total 16 trigger cards are needed to process 
the complete chamber information, distributed across four crates. 
The trigger card, realized as an 8-layer PCB, was designed in 
cooperation with University of Heidelberg
~\cite{ewhd}.
Each trigger card contains two Altera FPGAs~\cite{fpga}, 
mounted in a ball grid array technique onto the PCB. 
The board is equipped with a 5\,V to 1.8\,V power 
converter to deliver the 1.8\,V FPGA
kernel voltage. External 3.3\,V and 5\,V power supplies are needed for the 
FPGA I/O pins and the VME
bus and the IsPL~\cite{isplsi-1048} who controls it (see
Fig.\,\ref{triggercard}). Six EEPROMs~\cite{epc2-lc20}, three for every FPGA 
contain the configuration code, three for every FPGA. 

To transport the chamber information from the 250-pin input connector on 
the trigger card to the FPGAs 150\,lines are needed. 
The trigger algorithm is distributed over two FPGAs, because
one does not suffice for a complete
$\varphi$-sector. Every 
FPGA deals with the information of 60\,pads of five layers 
($15\times 4$ multiplexed channels). 
Tracks that point to pads of both CIPixes
define the overlap area, known to both FPGAs. 
A 90\,line interconnect bus connects both FPGAs. 
Information evaluated by FPGA\,I is transferred to FPGA\,II.
FPGA\,II is connected to two SCSI-connectors to deliver $\varphi$-based data 
to the sum cards via LVDS converters~\cite{ds90lv031}, which transform the 
3.3\,V FPGA output (LVTTL, low voltage TTL) to a LVDS (low voltage
differential standard) signal.  

Every trigger card is equipped with a 32-bit VME bus controller 
residing  in an in-system high density programmable logic device 
(IsPL)~\cite{isplsi-1048}. The 24-bit standard non-privileged address mode 
is used. 32 bidirectional data lines are linked to both FPGAs and to the
ISpL, which is set up as a 
VME-bus slave unit and organizes the data transfer from the FPGAs to the 
VME bus master.

The pad information of the chamber is stored in a ring buffer (pipeline) 
32 bunch crossings deep. Each trigger
card provides the information of one $\varphi$-sector. In case of a positive 
level\,1 trigger decision, data
taking stops and the readout of the triggered event is organized. Five bunch 
crossings around the expected
$ep$-event are read out. It takes 20 VME read cycles to read out one BC of one 
TC, 320 cycles per BC in total
\cite{phdurban}. 

\subsubsection{Summing crate}\label{sssec:sumcrate} 

The trigger information of each
trigger card is transferred to the summing crate 
(VME 21-slot 64x~\cite{wiener}). Here all 16 $\varphi$
sectors are combined and a trigger decision is calculated. Five 
identical sum cards are used to perform this task.
Four cards receive the information of four trigger crates each 
and combine it for a quadrant
(pre-sum card, see Fig.\ref{syst_scheme_trailer}). The fifth card, the 
main sum card, combines everything and distributes the decision 
(the trigger elements (TEs)) to the H1 central trigger control (CTC).
For monitoring, inputs and output signals of each sum card are read out. 

\begin{figure}[h]
\centering
\includegraphics[width=0.98\textwidth]{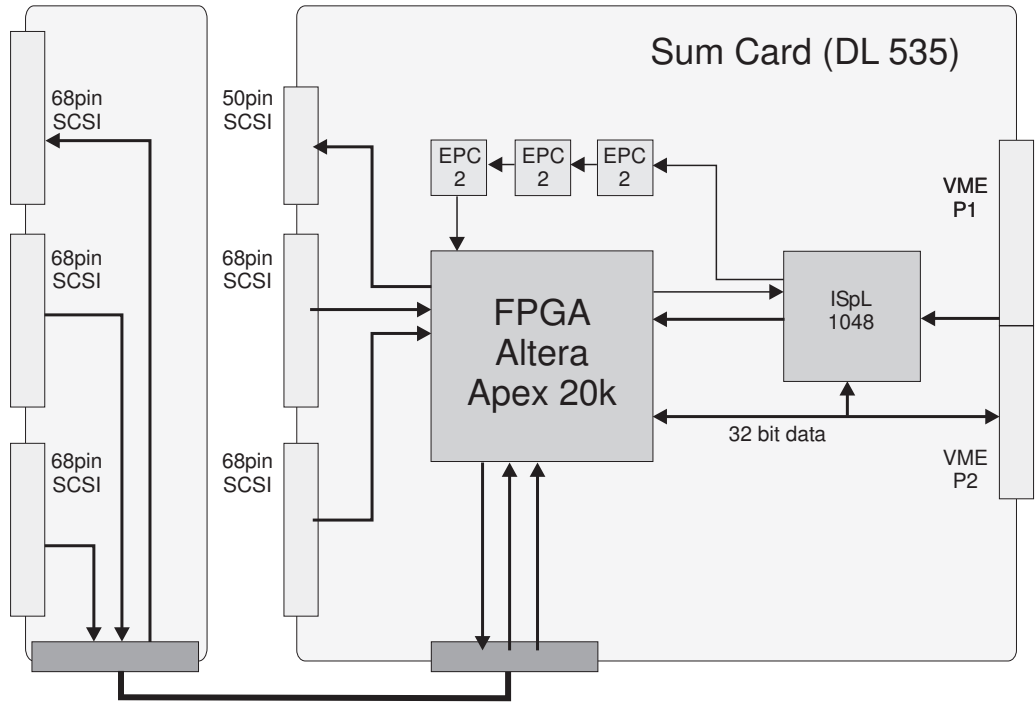}
\caption[The sum card]
{Schematic view of the sum card. Six SCSI connectors are connected to 
the FPGA. On the left side the piggy back
board is shown which provides three 68-pin SCSI connections. Three EPC2 
devices (EEPROMs) hold the configured data
in case of a power failure. A 32\,bit VME bus supports readout and 
programming of the FPGA (6 address bits). 
}\label{sumcard_scheme}
\end{figure}

The sum card was also developed in 
cooperation with University of Heidelberg~\cite{ewhd},
and is built around a single FPGA and an ISpL for configuration 
and VME bus service. The sum card is
equipped with a 32-bit VME bus (6 address bits in a 16-bit mode, AM 0x29). 
For the configuration of the FPGA,
three EEPROMs~\cite{epc2-lc20} are used.
Each sum card is equipped with six SCSI connectors 
(five 68-pin and one 50-pin micro sub-D). The signal levels meet the 
LVDS standard. Each 68-pin connector
transfers 32 single channels; the 50-pin connector transports 24 channels. In 
Figure\,\ref{sumcard_scheme} the sum card design is shown in a
schematic view.

Each sum card has a piggy back board to cope with the large number 
of signals, which are processed in the sum
card. Three 68-pin SCSI connectors are mounted on the piggy back board 
and are connected to the FPGA. The SCSI
connectors can be used as input or output channels, depending on the 
LVDS converter/driver chip~\cite{ds90lv031}. 
The master sum card receives the signals from the four pre-sum cards 
by four 68-pin input connectors,
the trigger decision is provided on the 50-pin output connector.

\subsubsection{Komposti crates}\label{sssec:komposti}

The {\em komposti} crate (VME 21-slot~\cite{wiener}) holds eight {\em komposti}
cards, which use the same hardware as the sum card, but differ 
in the programming of the FPGA. The
{\em komposti} card demultiplexes the data of the two innermost layers 
(120 pads of layer\,0 and 1 of 
two $\varphi$-sectors) and merges it
to 56 output signals provided at the two output SCSI connectors.
Four pads of the new chamber are merged to one input channel for the old 
trigger system~\cite{eic93}. 
The merged data is linked into 8 passive (without power supply and active 
electronic elements) converter cards
that form the LVDS signal of the {\em komposti} card to a pseudo-analog 
differential input signal in two 50-pin
connectors. Together with the pad information of the two layers of the outer 
proportional chamber (COP), the old $z$-vertex
trigger can be used in a 3-out-of-4 layer option \cite{eic93}, since the 
information of layer\,0 and layer\,1 is
combined at the {\em komposti} card (logical AND).

Running the old trigger in parallel was demanded because it entailed a
threshold cut on the transverse momentum of a track $p_t>0.5$ G$e$V/$c$, 
had a finer $z-$resolution, and it also served calorimeter trigger
towers.

\subsection{Data acquisition}\label{ssec:daq}

The CIP2k DAQ system consists of four VME CPU processor boards~\cite{CES}, 
which communicate with
the trigger cards, do the initial setup of the trigger-system and read 
out the stored chamber information from
the ring buffer in the FPGA. Each CPU reads the data of four $\varphi$ 
sectors. The CPU and the trigger card communicate 
via the VME bus, 
using single cycle D32/A24 access. 
The CPUs are interconnected by a PCI vertical interconnect bus 
system \cite{CES}.

Once the H1 central trigger decides to keep an event, all four CPUs start
the readout simultaneously, triggered by high priority interrupts.
The readout time is dominated by reading the pad data of
five time slices from the VME bus. Using the pVIC bus the information
from all $\varphi$ sectors is collected on a single crate. This synchronous
part of the readout ends after 1 ms, as required for H1 data acquisition
systems. 16 buffers are available at this stage, in order to avoid
possible second order deadtime effects caused by trigger bursts.
The asynchronous part of the readout operates on these buffers. It starts
with a data reduction reduction process 
on the main CPU. The chamber data, which are
stored as bit patterns, are translated into 16-bit wide pad numbers, resulting 
in an effective zero suppression if the occupancy of the chamber is less than 
10 \%.  During this process, pads that are known to be
bad are removed from the data. Identical filters are applied
during the trigger algorithm in order to ensure consistency of data and
trigger. The data are then transferred by the  VME bus to
the H1 central data acquisition 
system~\cite{becker1}.

\subsubsection{Control system}\label{ssec:control}

An $I^2C$-bus \cite{i2cbus} based system was developed to program 
the registers of the CIPix 
\cite{da_vollhardt}. The system is controlled by an 8-bit microprocessor 
(pic16F877) that is addressed via a
RS232 serial port connection. The programming of the CIPix settings is 
done by a Linux-based PC, installed in
the H1 control room with a 25\,m long RS232 cable connection to the 
$I^2C$-bus micro controller in the electronics trailer (CIPix-trailer-box). 

Moreover, a temperature control system was developed, using a (separate) 
one-wire-bus~\cite{onewire}, which is
connected to a temperature measurement chip on each CIPix board at the 
detector. This chip is connected to a second 
serial port of the CIPix-trailer-box and sends the measured
temperatures to the CPU in the {\em komposti} crate, 
The VME CPU receives the actual temperature and sends a continuous signal 
to a watch dog card. In case of high temperature or any other failure 
this signal stops and the watchdog switches off all CIPix
board power supplies.

 % ****************************************************************
 % CIP2k System Performance:  
% \input{nimperformance} 

\section{System performance}

The CIP2K trigger system is fully operational since autumn 2003. Some basic
performance tests of the new chamber and trigger electronics will be presented
in this section.
The comparisons presented below are all based on particle trajectories
reconstructed in the central jet chamber (CJC). The spatial resolution of the
CIP2K chamber and the trigger timing as well as single pad efficiencies
are studied with muons from cosmic rays, trigger efficiencies with $ep$ data.

\subsection{Spatial resolution}
\begin{figure}[htbp]
 \begin{center}
\parbox{150mm}{
  \includegraphics[width=70mm]{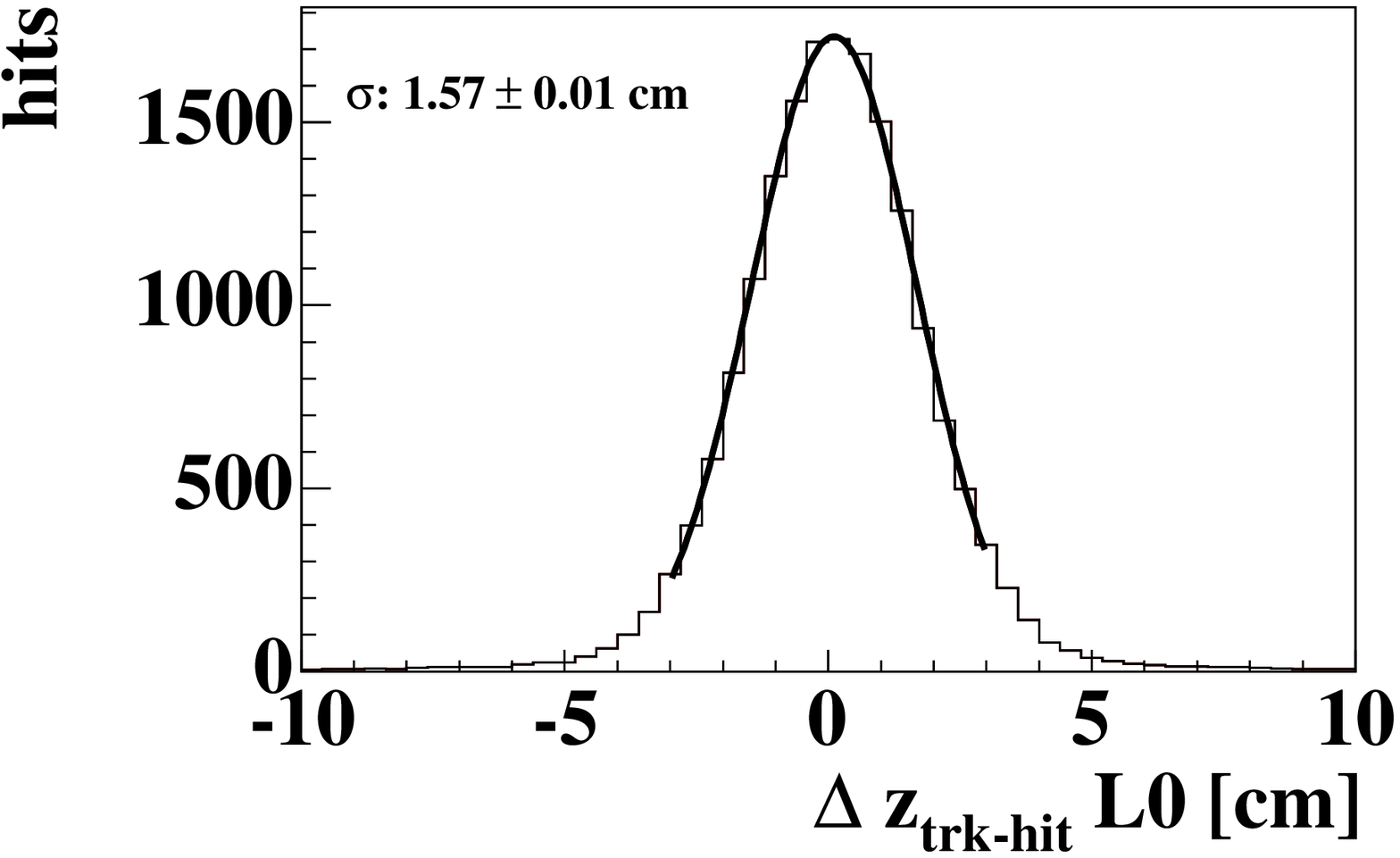}\hfill
  \includegraphics[width=70mm]{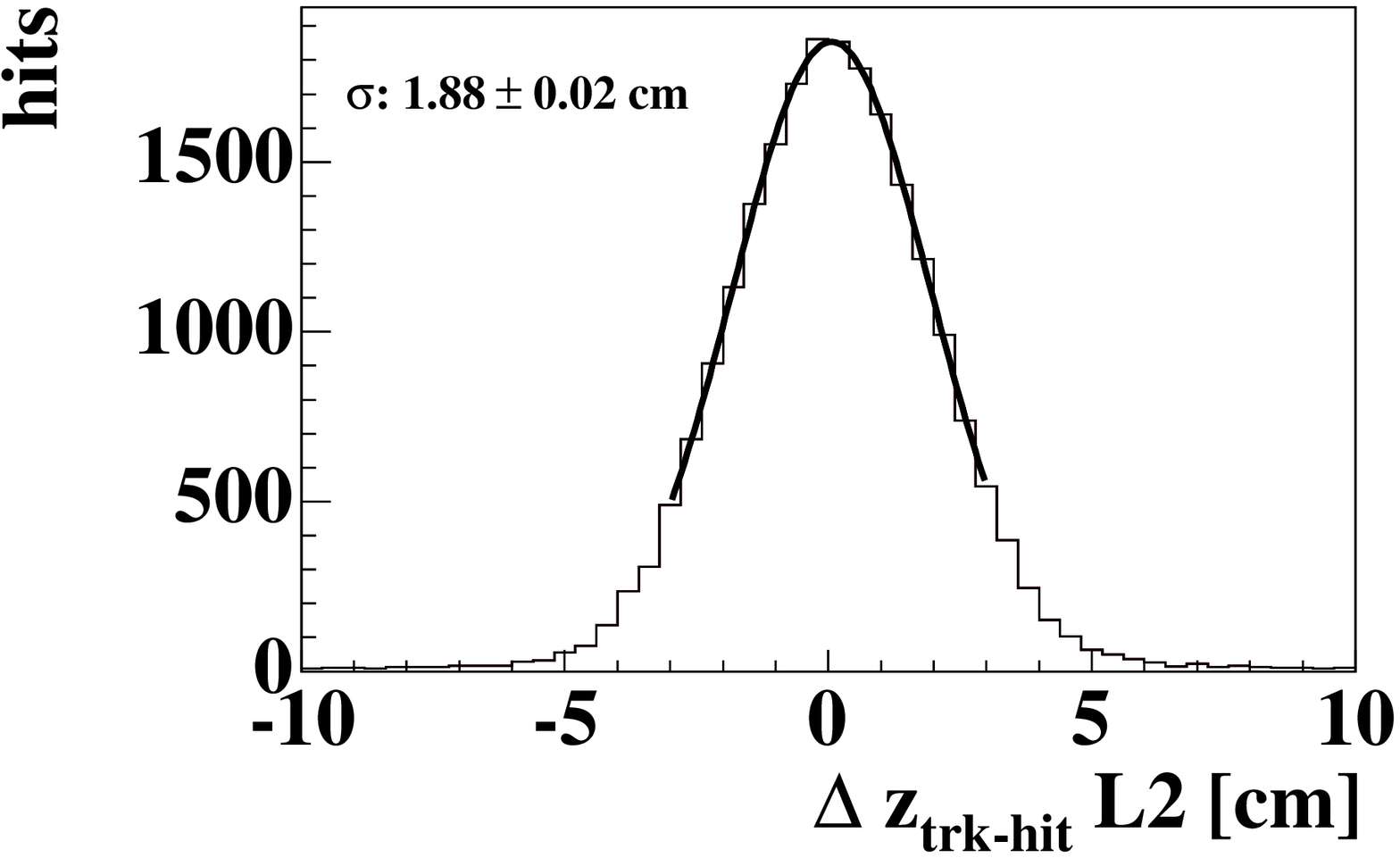}}
   \includegraphics[width=70mm]{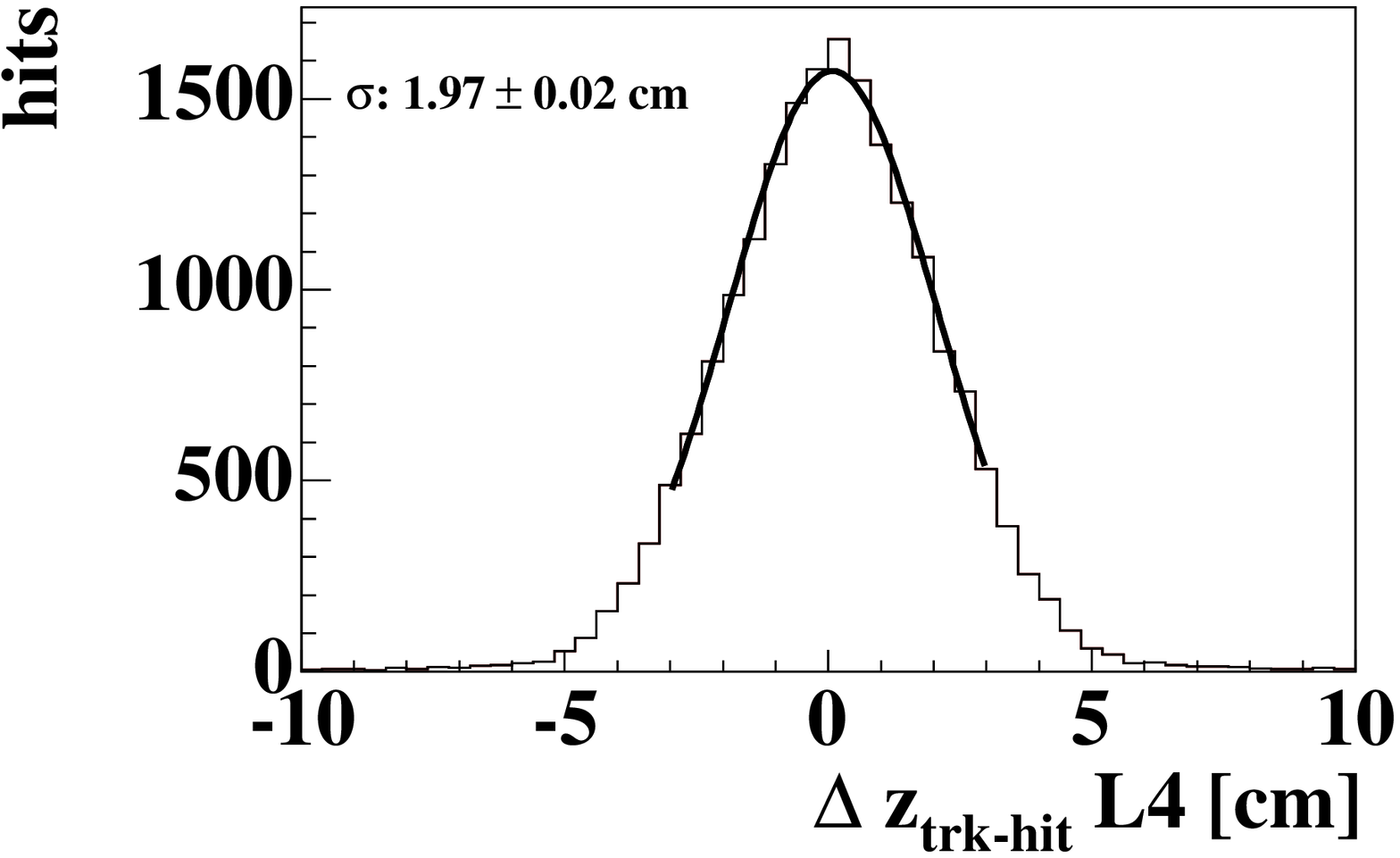}
 \end{center}
 \caption{The difference between the $z$-coordinate of drift-chamber tracks,
  extrapolated to the anode planes, and the centre position of active 
CIP2K pads of the corresponding azimuthal sector.}
\label{figpadres}
\end{figure}

Events with cosmic muons are selected to determine the spatial resolution.
 These events consist of two track halves, reconstructed in opposite
 sides of the CJC. The two track halves are constrained to have common track
 parameters, which improves the CJC $z$ position measurement
 significantly.  
 Figure \ref{figpadres} shows the difference in $z$ between the track 
position extrapolated to the CIP2K wire plane and the center of the 
active pads in the corresponding azimuthal sector for three layers.

The RMS-width of the distribution is $1.9\,\mathrm{cm}$, $2.2\,\mathrm{cm}$ 
and $2.3\,\mathrm{cm}$. This is in good agreement 
with the expected width, dominated by the finite pad size and the $z$ 
resolution 
of the CJC chamber. 
The chamber noise is negligible, as there are virtually no 
entries outside $\pm 6\,\mathrm{cm}$. 

\subsection{Trigger timing}
\begin{figure}[ht]
\centering
\includegraphics[width=0.4\textwidth]{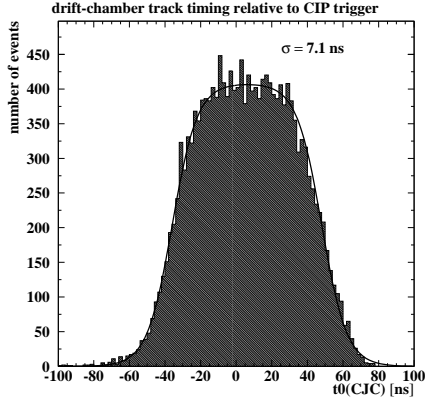}
\caption[Time resolution.]
{Time distribution of muons triggered by the CIP2K T0 trigger .}
\label{figtiming}
\end{figure}

The trigger timing is studied using the event sample with cosmic rays. 
Since the muons are not synchronized to the HERA beam clock it is possible 
to reconstruct the time resolution of the CIP2K 
system.
The events are triggered by the drift-chamber trigger~\cite{dcrphi}, 
which has only a moderate time
resolution ($\sigma_{\tau}\gg 50\,\mathrm{ns}$). 
 Figure
\ref{figtiming} shows the CJC event time for events with the CIP2K
T0 trigger active.
The CJC event time is the time at which the muon was closest to the $z$
axis. The time distribution is well described by the product of two
threshold functions with $\sigma$ the parameter for the intrinsic 
resolution of the CIP2K chambers and the electronics:
\[
f(t)=N\left[1+\exp(\frac{\mu_1-x}{\sigma})\right]^{-1}
\left[1+\exp(\frac{x-\mu_2}{\sigma})\right]^{-1}
\]
The width is found to be $\sigma=7.1\,\mathrm{ns}$, well compatible with
expectations from chamber and electronics design.

\subsection{Chamber and trigger efficiency}
\begin{figure}[htbp]
  \begin{center}
  \includegraphics[width=65mm]{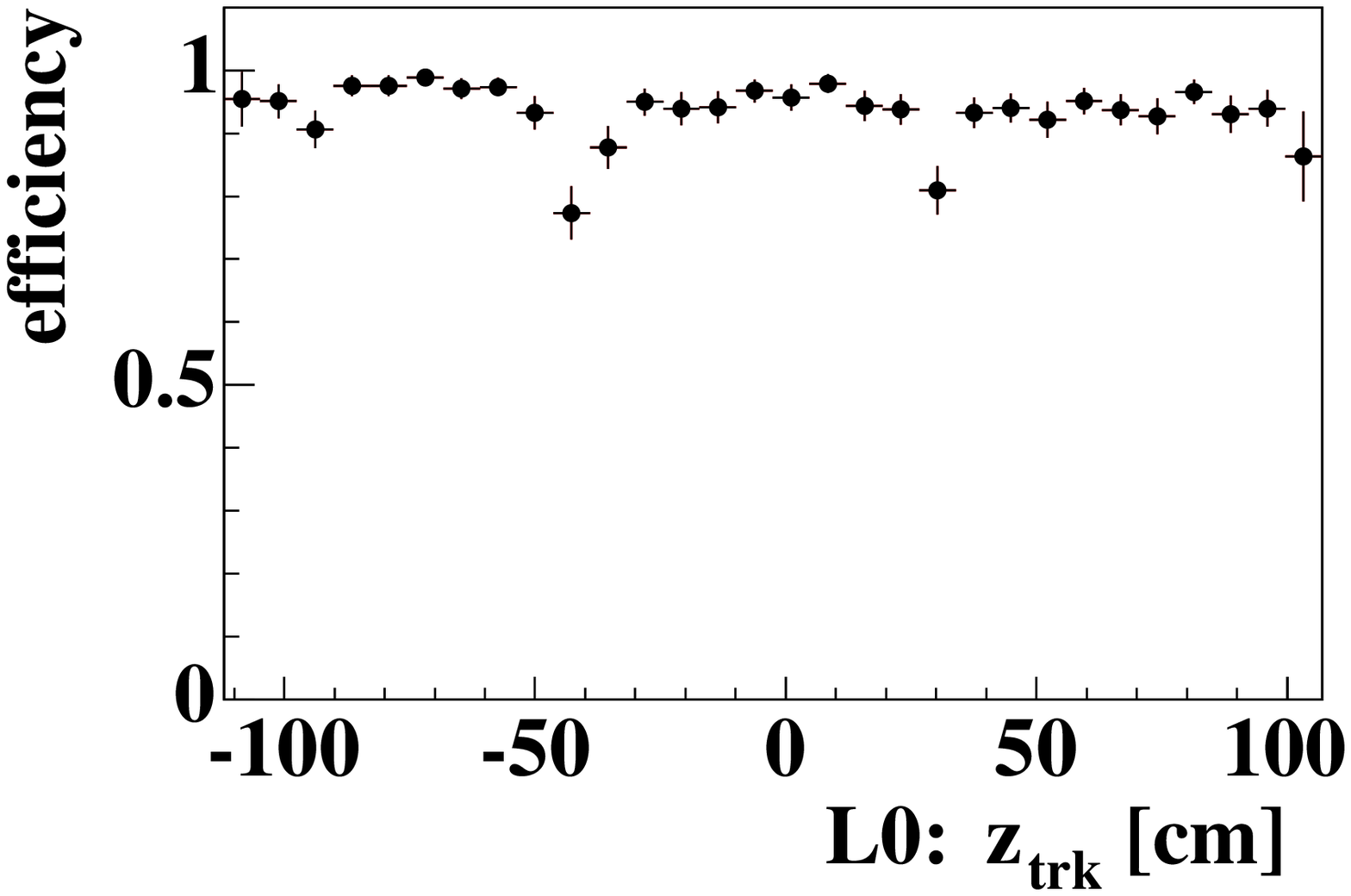} 
  \includegraphics[width=65mm]{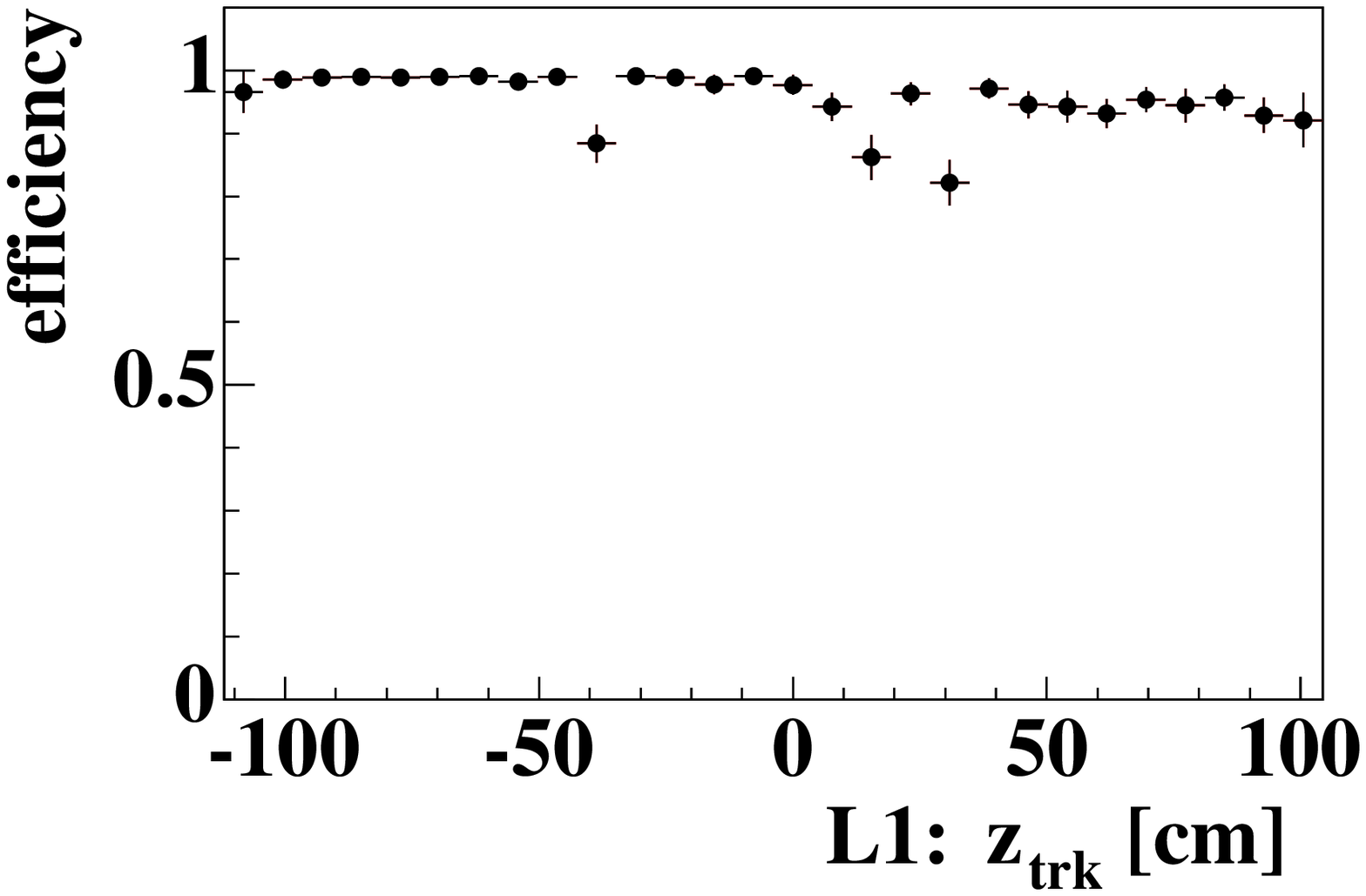} 
\includegraphics[width=65mm]{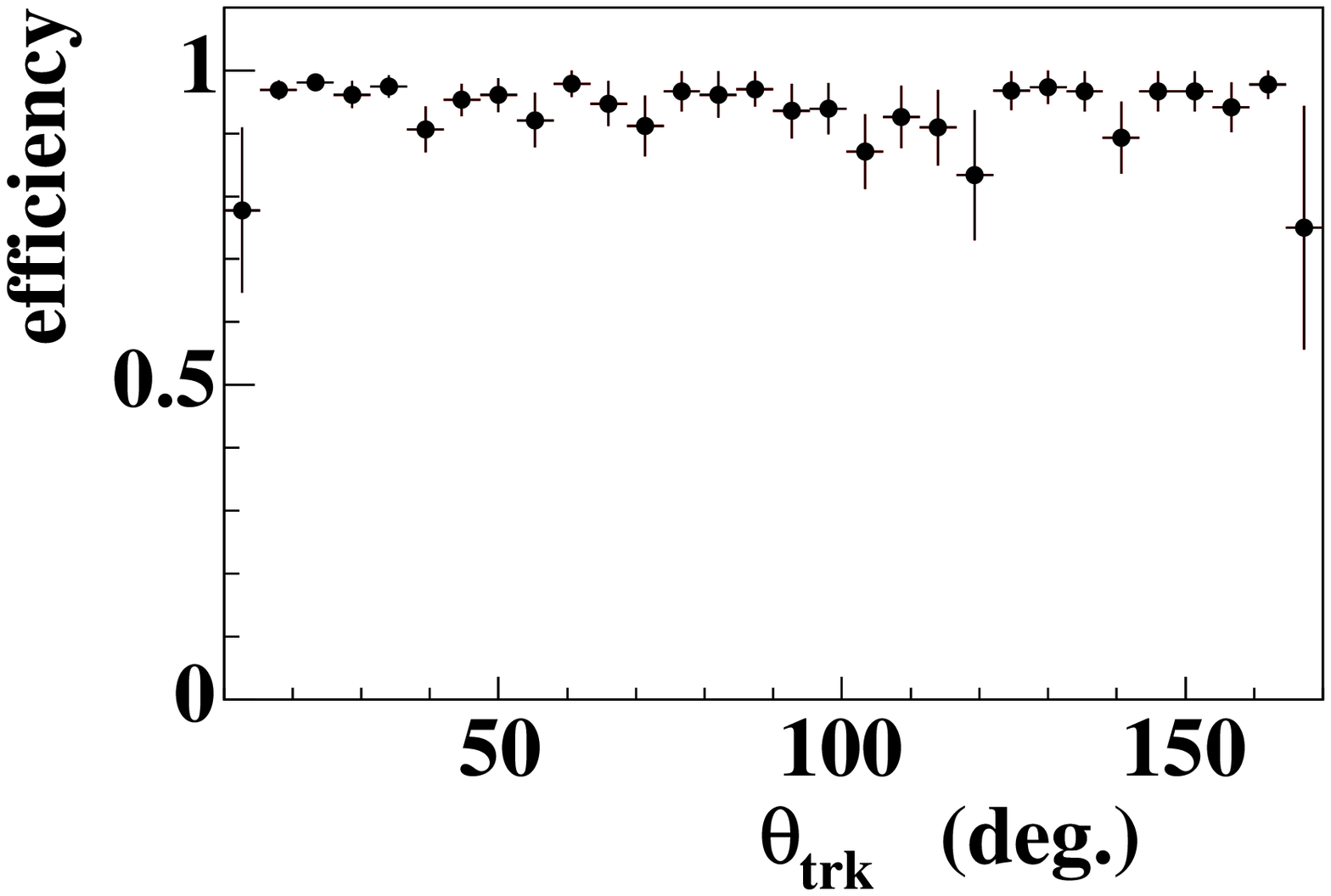}
\includegraphics[width=65mm]{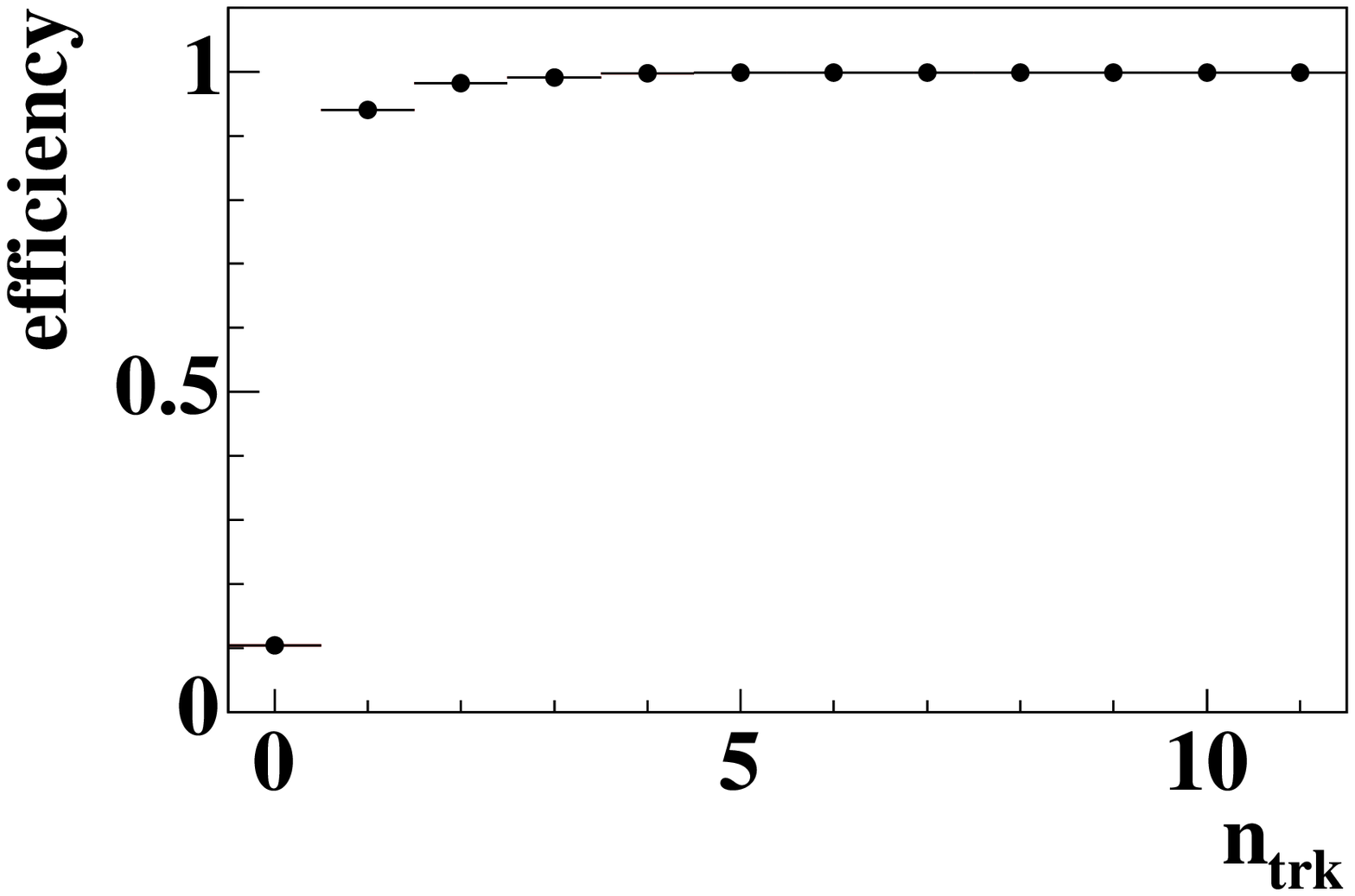}

 \end{center}
 \caption{Chamber efficiency for layer 0 (a) and 1 (b) as a function of 
the $z$-position of the tracks for cosmic muon events (top: left and right). 
(c) CIP2K trigger track 
efficiency as a function of the polar angle of the track, (d) $t_0$ 
efficiency as a function of the track multiplicity for  $ep$ data (bottom:
left and right). 
\label{figsingletrack}}
\end{figure}

The chamber efficiency was also deduced 
from the muon sample. 
The tracks in this sample are predominantly produced at polar
angles $\Theta\approx90^{\circ}$. They cover the full range along the $z$
axis. Figures~\ref{figsingletrack}(a) and (b) show the pad efficiency 
for all pads in one azimuthal sector in layer 0 and layer 1. 
Clearly visible is the reduced 
efficiency in both planes at $z\simeq-40$ and $z\simeq 33$~cm which 
corresponds to the position of the support rings for the anode wires.
Figure~\ref{figsingletrack}(c) shows the trigger track efficiency which is 
measured with $ep$ data as a function of the polar angle 
$\Theta$. For this study tracks 
are selected which are isolated in the CJC. A CIP2K trigger track requires 
a valid pattern with hits in at least four layers. 
The single-track trigger efficiency in
$ep$ events is  better than in the muon sample, because
most tracks in $ep$ events have polar angles different from $90^{\circ}$, so
their average energy loss is higher. Furthermore, tracks
passing the chamber at these polar angles may induce signals on several
adjacent pads, again increasing the detection efficiency. The wire support
structures are arranged far from the nominal interaction point, and their
impact on the detection efficiency of tracks from $ep$ collisions is not 
visible anymore.
Lastly  Fig.~\ref{figsingletrack}(d) shows the $t_0$ efficiency - 
the efficiency for finding a CIP2k track in the central region needed for
event timing - as a function of the track 
multiplicity. For events with only one reconstructed track it is 95 \%. This 
includes also inefficiencies due to broken readout electronics and HV 
problems.

\subsection{Background discrimination}
\begin{figure}[ht]
\centering
\includegraphics[width=0.8\textwidth]{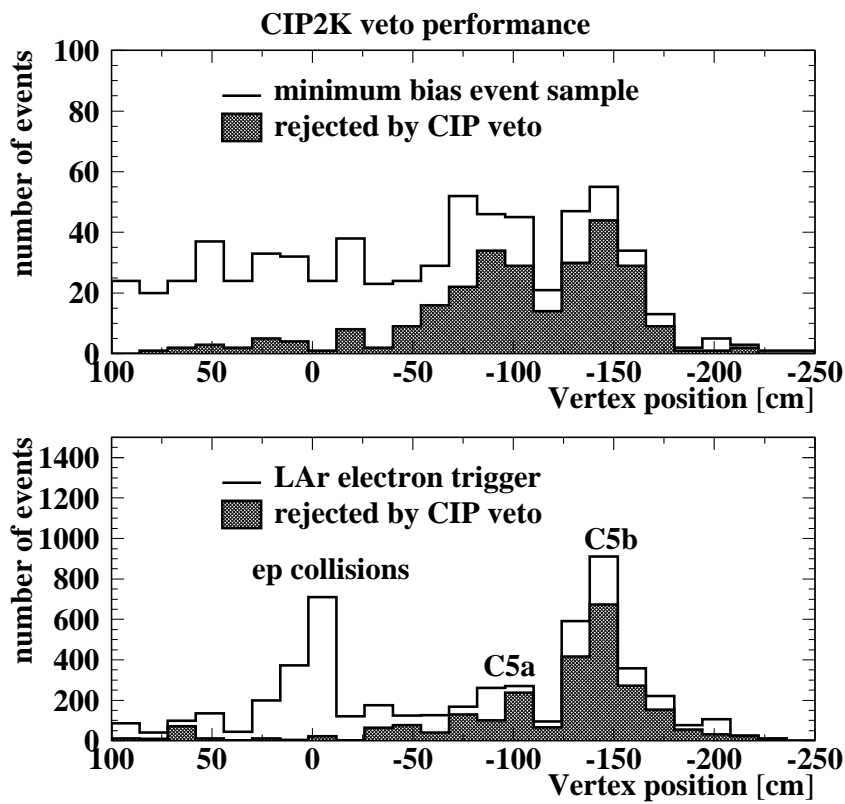}
\caption[Veto performance.]
{The $z$-vertex position based veto condition for two different event 
types. The shaded histogram shows the events
which are rejected by the CIP2K veto condition.}
\label{figveto}
\end{figure}

One of the main tasks of the CIP2K trigger system is the discrimination of
$ep$ events and high-multiplicity background events with displaced $z$-vertex
at the first trigger level. For many physics channels a veto condition is
applied on the first trigger level, based on CIP2K signals. An example of a
fairly soft veto condition is the following:
\begin{itemize}
\item more than 100 tracks found in the CIP2K system
\item the number of tracks with reconstructed $z$-vertex in the
  central region must not exceed the number of tracks with reconstructed
  $z$-vertex in the backward region.
\end{itemize}
Two different event samples have been studied to monitor the performance of 
this veto condition.
A minimum bias sample has been selected based on  randomly triggered
events, independent of any detector activity.
Alternatively, the veto condition was studied for events triggered
by a localized energy deposition in the liquid argon calorimeter (LAr electron
trigger).
Only those events are considered in the analysis with
at least one good CJC track within the CIP2K acceptance.
The CJC tracks are required to have a minimum of 20 detector hits,
in order to ensure a good $z$ coordinate measurement.
Since the $z$-vertex reconstruction often fails for background events,
the $z$-vertex position of the primary
interaction was estimated using the track with the largest number of hits.
Events with tracks not pointing back to the same vertex within
$250\,\mathrm{cm}$ were excluded from the analysis. This removes events where
the track reconstruction failed, but retains most events with multiple
interaction vertices.
Figure \ref{figveto} shows the $z$-vertex position as defined above for the 
minimum bias sample (top) and the sample
triggered by the LAR trigger (bottom). The shaded histogram shows the events 
which are rejected by the CIP2K veto condition.

The minimum bias sample is dominated by background events, such as 
collisions of protons or electrons with gas molecules in the beam pipe.
This is clearly visible in the vertex position histogram. The two peaks  near
$z=-90\,\mathrm{cm}$ and $Z=-150\,\mathrm{cm}$, correspond to the position of
the collimators C5a and C5b, respectively.
Secondary interactions of proton-gas interactions in the collimator produce
events with high particle multiplicities.
The flat distribution along $z$  mainly stems from electron-gas interactions.
These events usually have low track multiplicities.
No enhancement around the nominal interaction region at 
$z=0\pm 30~$cm is visible.
The background rejection by the $z$-vertex trigger can be further improved if 
the high multiplicity cut of the veto condition is reduced. In combination
with early signals from the backward scintillator veto wall~\cite{h1det} the 
rejection in the collimator region reaches values close to 100 \%.

%%% Local Variables: 
%%% mode: latex
%%% TeX-master: t
%%% End: 

 % ****************************************************************
 % Summary:   
% \input{nimsummary} 
\section{Summary}
\label {sec:summary}
This publication describes the new five
layer multi wire proportional chamber CIP2k with optical readout 
and a $z$-vertex trigger system for the H1 detector at HERA~II.
The main purpose of the trigger system is the
suppression of high multiplicity background events originating from the
backward part of the detector at the first trigger level.
After the luminosity upgrade of the HERA collider, the rate of such 
events has increased significantly. The new chamber with the $z$-vertex trigger
system is fully operational since autumn 2003. 
The trigger system is capable
of reconstructing the $z$-vertex position in parallel to the HERA beam clock
at $10.4\,\mathrm{MHz}$. The latency of the trigger decision is only
$1.5\,\mathrm{\mu s}$. 
This was achieved by choosing FPGA technology for the
track finding and histograming algorithms. 

The performance of the new CIP2K trigger system was analyzed. The chamber
behaves as expected in terms of spatial and time resolution. The single track
trigger-efficiencies are close to $100$ \%. The CIP2k trigger decision serves
as the main time base for other track based trigger systems. 
A veto condition for rejecting
background produced by secondary interactions in collimators in the vicinity
of the interaction point was studied in greater detail.
Such algorithms have become
an integral part of the H1 trigger system, and help to keep the H1 data
taking efficiency high.

 % ****************************************************************
 % Bibliography:                
% \input{nimbiblio}

%%% Local Variables: 
%%% mode: latex
%%% TeX-master: t
%%% End: 

\end{document}